\documentclass[aps,prl,twocolumn,superscriptaddress]{revtex4-1}
\usepackage{verbatim}
\usepackage{amsmath}
\usepackage[utf8]{inputenc}
\usepackage[pdftex]{graphicx}
\usepackage[separate-uncertainty=true]{siunitx}
\DeclareSIUnit{\rpm}{rpm}
\usepackage[colorlinks=true,urlcolor=blue,linkcolor=blue,citecolor=blue]{hyperref}
\usepackage{xcolor}
\usepackage[T1]{fontenc}
\usepackage{placeins}
\usepackage{nicefrac}
\usepackage{braket}
\usepackage{pdfcomment}
\usepackage{xcolor}
\usepackage{braket}
\usepackage{lineno}
\usepackage{atbegshi}

\newcommand{\HS}{MoS$_2$/WS$_2$}

\newcommand{\HSe}{MoSe$_2$/WSe$_2$}

\newcommand{\moly}{MoS$_2$}

\newcommand{\woly}{WS$_2$}
\newcommand{\mole}{MoSe$_2$}
\newcommand{\wole}{WSe$_2$}
\newcommand{\percent}[1]{\SI{#1}{\percent}}
\renewcommand{\deg}{^\circ}

\begin{document}

\date{\today}

\author{Thorsten Deilmann}
\affiliation{Institut für Festkörertheorie, Westfälische Wilhelms-Universität Münster, Wilhelm-Klemm-Str.10, 48149 Münster, Germany}

\author{Michael Rohlfing}
\affiliation{Institut für Festkörertheorie, Westfälische Wilhelms-Universität Münster, Wilhelm-Klemm-Str.10, 48149 Münster, Germany}

\author{Ursula Wurstbauer}\email{wurstbauer@wwu.de}
\affiliation{Institute of Physics, Westfälische Wilhelms-Universität Münster, Wilhelm-Klemm-Str.10, 48149 Münster, Germany}

\title{Light-matter interaction in van der Waals heterostructures}



\begin{abstract}
  Even if individual two-dimensional materials own various interesting and unexpected properties,
  the stacking of such layers leads to van der Waals solids
  which unite the characteristics of two dimensions
  with novel features originating from the interlayer interactions.
  In this topical review, we cover fabrication and characterization of van der Waals heterosructures with a focus on hetero-bilayers made of monolayers of semiconducting transition metal dichalcogenides.
  Experimental and theoretical techniques to investigate those heterobilayers are introduced.
  Most recent findings focusing on different
  transition metal dichalcogenides heterostructures are presented and possible optical transitions between different valleys, appearance of moiré patterns and signatures of moiré excitons are discussed. The fascinating and fast growing research on van der Waals hetero-bilayers provide promising insights required for their application as emerging quantum-nano materials.
\end{abstract}


\maketitle
\section{Introduction}\label{sec:intro}
\begin{figure}
\includegraphics[scale=1]{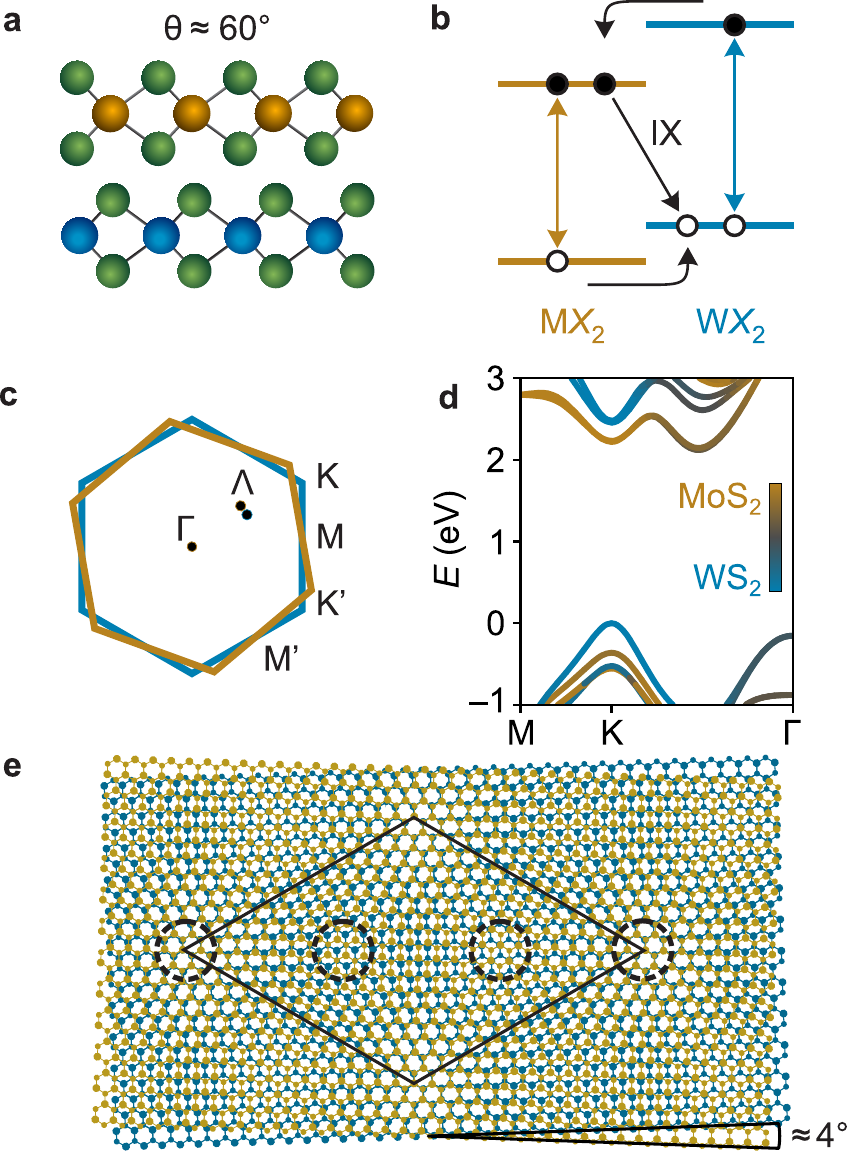}
\caption{a) Side view of a ball representation of TMDC monolayer crystals AB stacked into a hetero-bilayer (twist angle of $60\deg$).
b) Sketch of the idealized staggered type-II band alignment of TMDC hetero-bilayers resulting in an efficient charge transfer between the layers of photo-generated electron hole pairs and the formation of interlayer excitons (IXs) with reduced transition energy.
c) Overlapped first Brillouin zones of two layers with finite mismatch of the lattice constants and a small deviation of the rotational alignment from AA or AB stacking resulting in a displacement, e.g., of the K / K' and the $\lambda$ points.
d) Ab initio calculation of the lowest conduction bands and topmost valence bands of the lattice-matched \moly/\woly\ hetero-bilayer based on density functional theory and many-body perturbation theory. The colors indicate the contributions of the individual layers as well as their mixture.
The energy scale has been aligned to the valence band maximum.
e) Nearly AB stacked lattices with identical lattice constants and a twist of about $4\deg$. The black diamond indicates the unit cell of the moiré pattern and the dashed circles indicate the regions, where the local arrangement of atoms exhibit 3-fold rotation symmetry ($C_{3}$ symmetry).
}
\label{fig: 1}
\end{figure}

Two-dimensional (2D) crystals are highly ordered covalently or ionically bonded crystals with a sheet thickness of only one or few atoms resulting in a thickness of typically less than 1 nm. Their parental materials are layered crystals in which the individual layers are coupled by weak van der Waals (vdW) forces. Only 15 years ago, the cleavage of graphite into single atomic layers, called graphene \cite{Novoselov.2004} started a new research area on 2D materials, also called vdW materials. After the discovery of graphene, the most prominent representative of this class of materials, many other 2D crystals have been identified, often with intriguing properties that have no counterparts in three-dimensional solids. 2D materials cover a wealth of physical, mechanical and chemical properties. The real advantage of those materials is that they can be arbitrarily combined by mechanical stacking without the constraints of the out-of-plane registry as for conventional 3D solids \cite{Novoselov.2005, Ajayan.2016} facilitating new device architectures. Due to proximity and hybridization effects, the assembly in such new ‘artificial vdW solids’ can result in the emergence of new states of matter with novel functionalities \cite{Novoselov.2016} that not only depend on the combination of different 2D crystal but also on their rotational alignment opening the avenue for a new field called \textit{twisttronics} \cite{Zhang.2017, Kang.2013b, RibeiroPalau.2018, Cao.2018, Hunt.2013, Dean.2013, Tran.2019, Wu.2017, Wu.2018, Seyler.2019} as indicated in figure \ref{fig: 1}c-e.

Since the first experimental evidence of a direct band gap in monolayer \moly\ in 2010 \cite{Mak.2010, Splendiani.2010}, a vivid and divers research area has evolved around transition metal dichalcogenides (TMDCs), which show a row of fascinating phenomena \cite{Wilson.1969, Mak.2016, Mueller.2018}. There are semiconducting, metallic, and superconducting systems among the TMDCs \cite{Wilson.1969, Miro.2014}. Semiconducting TMDCs such as \moly, \woly, \mole, and \wole\ are a class of 2D materials with unique optical \cite{Koperski.2017, Schneider.2018}, optoelectronic \cite{Mueller.2018, Parzinger.2017} and electronic \cite{Radisavljevic.2011} properties including exciton dominated optical response even at room temperature \cite{Chernikov.2014, He.2014, Wurstbauer.2017}, catalytic activity and stability \cite{Parzinger.2015, He.2016, Mitterreiter2019} doping induced superconductivity, fascinating spin and valley properties up to room temperature \cite{DiXiao.2012, Kormanyos.2014, Xu.2014, Mak.2018, Miller.2019}. The all-2D surface nature of TMDCs enables the activation of quantum-like emitters via strain or defect activation \cite{Blauth.2017, Tonndorf.2015, Srivastava.2015, Koperski.2015, Chakraborty.2015, Kern.2016, PalaciosBerraquero.2017, Branny.2017, Klein.2019}.

Van der Waals hetero-bilayers (HBs) prepared from optically active semiconducting TMDC monolayers, such as \moly\ or \woly, combine the excellent properties of the individual layers with the potential for novel functionalities provided by the full control of vdW architecture \cite{Geim.2013, Rivera.2018}.
Stacking of two different TMDC monolayers on top of each other, the staggered electronic bands create an atomically sharp p-n heterojunction, effectively separating photo-generated electron-hole pairs as sketched in figures \ref{fig: 1}a, b. This charge transfer \cite{Rigosi.2015, Zhu.2015b, Chen.2016, Rivera.2018, Mak.2018} allows for the formation of so-called interlayer excitons (IXs) a.k.a. composite bosons with electrons and holes residing in different layers \cite{Rivera.2015, Rivera.2016, Rivera.2018, Mak.2018}. The IX share the high binding energies and intriguing spin and valley properties with direct excitons in TMDC monolayers \cite{Wilson.2017, Merkl.2019}. The separation creates a permanent out-of plane dipolar moment facilitating the manipulation of IX with electric fields in gated structures \cite{Ciarrocchi.2018, Kiemle.2019}. The reduced overlap of the electron and hole wavefunctions results in long exciton lifetimes of more than \SI{100}{\nano\second} \cite{Miller.2017, Rivera.2015} greatly exceeding the lifetime of intralayer excitons \cite{Robert.2016, Palummo.2015}. Long lifetimes allow the thermalization of IX to lattice temperature. Diffusion of thermalized IX over several micrometers \cite{Unuchek.2018, Kulig.2018} and manipulation by electric fields allows the generation of dense exciton ensembles \cite{Nagler.2017} as well as the operation of functional excitonic devices \cite{Unuchek.2018, Jauregui.2018}. This synergy of fascinating physical properties makes the TMDC HBs not only a promising platform for application in the area of solid state lighting \cite{Withers.2015}, energy conversion \cite{Parzinger.2017}, as well as opto- and valleytronics \cite{Jones.2013, Schaibley.2016}, but also in exciton based information technologies \cite{Unuchek.2018, Ciarrocchi.2018}, and in the design of novel designer quantum nanomaterials. Moreover, vdW structures exhibit potential to access collective excitonic phenomena like high-temperature superfluidity and Bose-Einstein condensation\cite{Fogler.2014, Wang.2019, Sigl.2020} as well as excitons in moiré-superlattices \cite{Seyler.2019, Jin.2019, Tran.2019, Zhang.2018, Jin.2019b}.

Here, we review the fascinating light interaction properties with a focus on IX in semiconducting TMDC hetero-bilayers. We start with selected fabrication and characterization methods with a focus on the determination of the interlayer coupling strength and the precise measurement rotation angle between the individual layers using second harmonic generation (SHG) and Raman spectroscopy. Next we review theoretical approaches to the optical properties of TMDC heterostructures. In section~\ref{sec:multi} we discuss the impact of multi-valley properties and hybridization on the formation resulting in rich IX multiplet emission with interesting properties. From these considerations we will see that the picture of a formation of an atomically sharp p-n junction is oversimplified for the explanation of the optical properties of TMDC hetero-bilayers. These vdW structures behaves more like newly created vdW solid than a lose heterostructure from two individual layers. These is even more evident in the next chapter~\ref{sec:moire}, where the influence of moiré potentials given by small lattice mismatch and / or twist between the two adjacent layers results in the formation of so called moiré excitons. This topical review is closed by a summary and a brief outlook on the potential of van der Waals heterostructures not only for electronic, optoelectronic, sensing and (solar) energy conversion applications, but also as promising quantum-nano materials supporting future key technologies.

\section{Fabrication and Characterization}\label{sec:fab}

\begin{figure}
\includegraphics[scale=1]{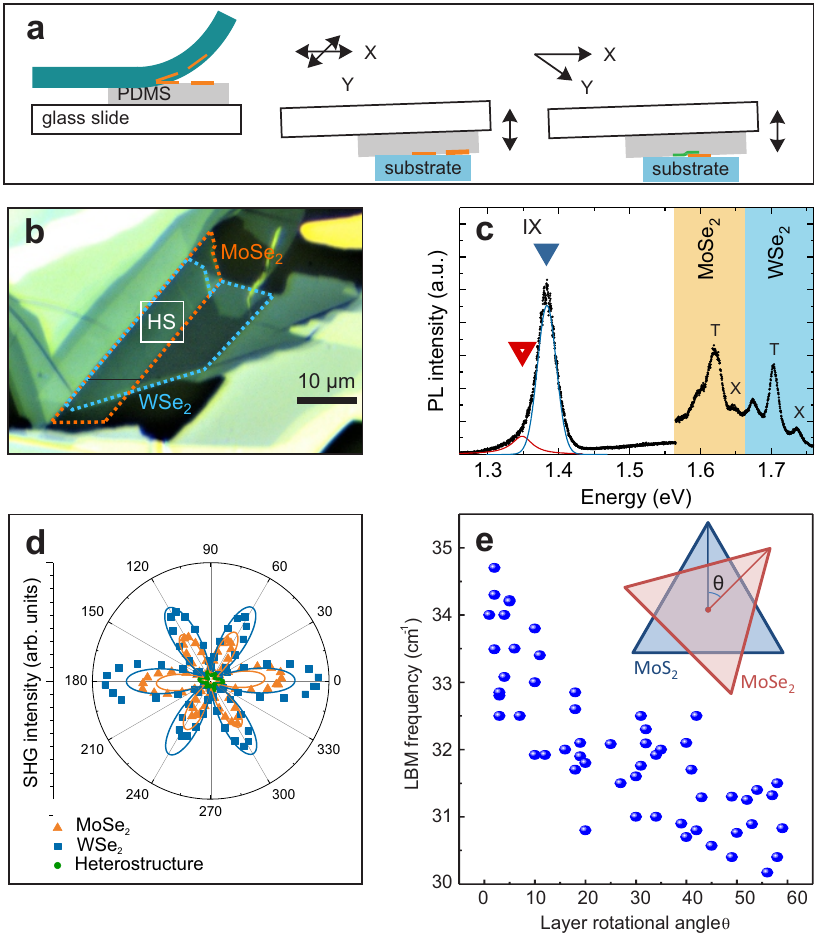}
\caption{a) Sketch of the most important steps during fabrication of vdW heterostructures by micromechanical exfoliation and vdW assembly. A thin bulk crystal on adhesive tape is mechanically exfoliated on a viscoelastic PDMS stamp, monolayer flakes are identified via optical microscopy aligned and transferred onto the target substrate. Same approach for the next layer.
b) Optical micrograph of a \HSe\ heterostructures consisting of a hetero-bilayer region (marked region in the center) \cite{Miller.2017}.
c) Low-temperature (3K) PL measurement on the \HSe\ hetero-bilayer shown in (b). Quenching of the direct excitons and a clear red-shifted signal from IX transition proofs strong interlayer coupling. The asymmetric IX peak indicates emission from an IX multiplet \cite{Miller.2017}.
[b,c) Reprinted with permission from ACS \cite{Miller.2017}.]
d) SHG intensity of isolated monolayers of \mole\ (orange triangulars) and \wole\ (blue squares) and the hetero-bilayer region (green circles). The intensity of the SHG signal is plotted versus the rotation angle $\theta$ of the parallel polarizers for incident and detected light.
e) Energy of the layer breathing phonon modes (LBM) from low-energy Raman spectroscopy in dependence of the twist angle $\theta$ for \moly/\mole\ hetero-bilayers \cite{Lui.2015}.
[(e) Reprinted with permission from \cite{Lui.2015}.]
}
\label{fig: 2}
\end{figure}

To exploit the exciting properties of vdW heterostructures, the synthesis of high-quality vdW crystals is essential. There are three routes for the synthesis of vdW structures \cite{lv.2015}: (a)  Direct vdW growth on flat inert substrates by powder based or gas source chemical vapor deposition (CVD) or molecular beam epitaxy (MBE). Direct growth of vdW heterostructure is challenging due to issues with nucleation sites resulting in small lateral domains, due to challenges in the precise control of the number of layers over large lateral dimensions and difficulties with intermixing and alloying hampering the formation of atomically sharp interfaces \cite{Briggs.2019}. (b)  Liquid exfoliation using either an intercalation approach or just application of mechanical forces by sonication with subsequent centrifugation to select crystallites of a certain thickness and size in solution \cite{Nicolosi.2013, Backes.2017}. This “ink” containing 2D crystallites with a typical size of a few micrometers can then be used for ink-printing spray or dip coating of a thin network of 2D crystals resulting in 2D networks, a method that is inapplicable to prepare at least several ten micrometer large mono-crystalline TMDC hetero-bilayers with high quality and atomically sharp interfaces. (c) Micromechanical exfoliation from bulk crystal or pick-up of CVD grown monolayers using adhesive tapes and the subsequent vdW assembly into heterostacks by various deterministic transfer processes \cite{CastellanosGomez.2014, Frisenda.2017, Masubuchi.2018, Pizzocchero.2016} as schematically illustrated in figure \ref{fig: 2}a. With help of mechanical translations- and rotations-stages for precise alignment, high quality samples with various combination of different materials and arbitrary number of layers can be assembled (figure \ref{fig: 2}b) e.g. combining different TMDC monolayers, graphene and hexagonal boron nitride (hBN) into field-effect structures \cite{Kiemle.2019}. The vdW assembly can even be performed in inert-gas atmosphere in a glove-box. This might be of importance for several applications since the properties of the heterostacks sensitively rely on the quality of the atomistic interfaces \cite{Rooney.2017}. Intriguingly, the vdW force between adjacent layers is for a variety of 2D material combination sufficient to accumulate contaminants into isolated bubbles, leaving large lateral regions with atomically clean interfaces in between \cite{Haigh.2012}, enabling very good interface quality even when doing the whole vdW assembly process in ambient.

The vertical distance between adjacent layers and their rotational alignment determines the strength of the interlayer interaction that in turn quantifies the extent to which the properties of the heterostructure deviate from the combined properties of individual isolated layers. The effect of interlayer interaction is reflected in the electronic band structure of TMDC hetero-bilayers and a sufficient degree of interlayer interaction is required for efficient charge transfer between the adjacent layers essential for the formation of IX. Under optical excitation above the band gap of the constituent TMDC monolayers, electron hole pairs are preferably generated within one layer due to the small oscillator strength of the interband absorption at the IX transition \cite{Komsa.2013} as illustrated in figure \ref{fig: 1}b. Ultrafast separation of electrons and holes across the heterojunction \cite{Hong.2014, Ceballos.2014, Rigosi.2015, Merkl.2019} results in accumulation of electrons in one layer and holes in the other layer, respectively. Strong delocalization \cite{Long.2016} and hybridized $\Gamma$ and $\Lambda$ valleys (c.f figure \ref{fig: 1}d) in the hetero-bilayers \cite{Wang.2017c} are discussed as possible reason for efficient charge separation. A clear signature for strong interlayer coupling of a TMDC heterobilayer is therefore a significantly quenched photoluminescence (PL) signal of the direct excitons compared to the corresponding isolated monolayers (c.f figure \ref{fig: 2}c) \cite{Chiu.2014, Fang.2014, Hong.2014, Miller.2017}. As shown in figure \ref{fig: 2}c) on the example of a \HSe hetero-bilayer, quenching of the intralayer exciton PL \cite{Tongay.2014, Alexeev.2017} and the appearance of a redshifted IX PL signal absent for the individual monolayers are commonly used as a measure for the strength of the interlayer coupling. The homogeneous broadening of linewidth of intra- and interlayer exciton PL can be greatly suppressed by hBN encapsulating the TMDC mono- and hetero-bilayers \cite{Cadiz.2017, Wierzbowski.2017, Ajayi.2017, Okada.2018, Ciarrocchi.2018, Kiemle.2019} enabling the observation of fine-structures in the (IX) emission energy as a result of e.g. formation of charged excitons (trions) \cite{Calman.2019}, spin-split bands \cite{Rivera.2015, Ciarrocchi.2018, Hanbicki.2018}, formation of momentum direct and indirect multi-valley excitons \cite{Miller.2017, Baranowski.2017, Kunstmann.2018, Okada.2018, Kiemle.2019} or moiré excitons \cite{Jin.2019, Jin.2019b, Seyler.2019, Tran.2019, Alexeev.2019}. Moreover, exciton diffussion over micrometers has been demonstrated for IX hosted in a \mole/\wole\ hetero-bilayer \cite{Jauregui.2018} making those systems promising candidates for the realization of excitonic devices \cite{Unuchek.2018} and to study many-body of thermalized excitonic ensembles.

Control of the twist angle between the stacked TMDC monolayers is essential since interlayer coupling, hybridization and moiré physics crucially depend on their rotational alignment. The alignment of the crystal axes during the transfer process with an accuracy of better than $\pm 1 - 2\deg$ can be achieved by determining and orienting the crystal axes by optical microscopy. This is possible since the crystal axes of the micromechanically exfoliated flakes breaks preferably under modulo $60\deg$. With this method, it is not possible to distinguish between AA ($0\deg$) and AB ($60\deg$) stacking order and it would be desired to know the actual twist angle with better precision. An accurate measurement of the stacking order and twist angle between adjacent layer is facilitated by polarization resolved second harmonic generation (SHG) experiments. Non linear terms of the electric susceptibility $\chi$ can cause the generation of higher harmonics of a strong incident light field \cite{Franken.1961}. The electric susceptibility $\chi$ is a tensor and reflects the crystal symmetry and in particular, the polarization induced by the second order of the susceptibility $\chi^{2}$ is non-zero only for non-centrosymmetric crystals as are TMDC monolayers and in more general TMDCs with an odd number of layers \cite{Li.2013, Kumar.2013, Kleemann.2017}. The crystal symmetry of TMDC monolayers dictates the intensity of the SHG $I(\chi^{2})$ to follow a $60\deg$ oscillation $I(\chi^{2}) \propto \sin^{2}(3\theta)$ with the angle $\theta$ between polarization vector of light and crystal axis \cite{Li.2013, Malard.2013}. Determination the SHG intensity $I(\theta)$ in dependence of the twist $\theta$ of the individual layers provides a precise measurement of the twist between the two layers disregarding AA or AB stacking. The SHG signal taken on a coupled TMDC hetero-bilayer results as for natural TMDC bilayers from interference of the second harmonic fields of the two constituent layers \cite{Hsu.2014}. The second harmonic fields of the hetero-bilayer interfere constructively for AA stacking and destructively for AB stacking order. Figure \ref{fig: 2}d displays SHG intensities in a polar representation of a \HSe\ hetero-bilayer indicating a twist between the two layers of about $\pm 2\deg$. The vanishing signal from the heterostructure region indicates AB stacking order \cite{Miller.2017}.

Another fingerprint for strong interlayer coupling is the observation of layer breathing phonon modes in TMDC hetero-bilayers by means of ultra-low frequency Raman spectroscopy \cite{Lui.2015}. The mode energies are between 30 cm$^{-1}$ and 40 cm$^{-1}$ and evolve systematically with the twist angle as shown in figure \ref{fig: 2}e on example of \moly/\mole\ hetero-bilayers \cite{Lui.2015}.

\section{Theoretical description of Intra- and Intervalley Excitons}\label{sec:theo}
In this chapter, we will discuss the theoretical description of
electronic and optical properties from first principles.
In the last years it has become clear that
monolayer as well as homogeneous and lattice-matched heterogeneous multilayer
can be described well by the commonly used $GW$-BSE
methodology and its extension to trions
which we will recapitulate in Sec.~\ref{theo:method}.
Thereafter, we will briefly discuss the ``universal'' properties of intralayer excitons (Sec.~\ref{theo:intra}) and,
finally, we highlight the possibilities
of excitons which are (partially) located on neighbouring layer
and are thus called interlayer states (Sec.~\ref{theo:inter}).

\subsection{Theoretical methodology: Ab-initio $GW$-BSE calculations}\label{theo:method}
A precise description of the optical and, in particular, of the excitonic properties
requires reliable structural and electronic properties as a starting point.
Geometric structures, especially chemical bonds (i.e. intralayer properties),
are often well described in the parameter-free density functional theory (DFT).
On the other hand, DFT is not sufficient for a quantitative description
of the electronic properties, e.g. the band gap, and more accurate methods are required.
In recent years many-body perturbation theory \cite{Rohlfing_eh, Onida_2002, Thygesen2D, TrionCNT}
(i.e. the $GW$ approximation for electronic properties,
the Bethe-Salpeter equation (BSE) for optical properties of neutral excitations,
and its extension for trions, i.e. charged excitations)
has turned out as promising ab-initio description.
The (quasi-particle) energies $E_{n,{\bf k}}$ and wave functions $\psi_{n,{\bf k}}$
can be calculated by evaluating the Dyson equation
\begin{equation}\label{eq:GW}
  \left[ H^\text{DFT} - V_\text{xc} + \Sigma(E_{n{\bf k}}) \right] \psi_{n{\bf k}} = E_{n{\bf k}} \psi_{n{\bf k}},
\end{equation}
in which $\Sigma$ is the self-energy \cite{HedinGW}
that supersedes the exchange correlation potential $V_\text{xc}$ obtained in DFT.
Based on this, the BSE can be setup as
\begin{equation}\label{eq:BSE}
  (E_{c{\bf k}} - E_{v{\bf k}}) A^{S}_{vc{\bf k}} + \sum_{v'c'{\bf k}'} K^\text{eh}_{vc{\bf k},v'c'{\bf k}'} A^{S}_{v'c'{\bf k}'} = \Omega^S A^S_{vc{\bf k}}~,
\end{equation}
where $K^\text{eh}$ is the direct and exchange electron-hole interaction
and $\Omega^S$/$A^{S}_{vc{\bf k}}$ denote the exciton energies and amplitudes.
$v$, $c$, and ${\bf k}$ count the hole states, electron states, and $k$ points, respectively.
For three-particle excitations Eq.~(\ref{eq:BSE}) can be extended to incorporate
the quasi-particle energy of a third particle and its additional electron-hole attraction as well as electron-electron repulsion \cite{TrionCNT}.

We underline that all three methods allow for system-independent and parameter-free descriptions.
In contrast to DFT which is able to describe the interactions mediated by the wave-function overlap very well,
the self-energy also incorporates long range non-local electrodynamical effects.
Thus, in particular, the influence of neighbouring layers is well described
and both intra- and interlayer excitations can be compared on equal footing.

In addition several parameter-driven models have been proposed to model two-dimensional materials \cite{chaves_theoretical_2016, Ganchev_2015, mostaani_diffusion_2017, Berkelbach.2013, Chernikov.2014, Selig_2016}.
Most of them rely on an effective mass picture for electrons and holes and the Rytova-Keldysh potential for the interaction between the particles.
While these models have been employed frequently for monolayers,
their usage for heterostructures is much more challenging.
The number of parameters for the atom- and distance-depending interlayer couplings
often prohibits simple and realistic models.
For the sake of brevity, we will focus on the description from first principles in this review.

\subsection{Intralayer excitons}\label{theo:intra}
The optical properties of monolayers,
such as \textit{h}BN, black phosphorus, and especially of TMDCs have been widely discussed in literature,
see  e.g. Refs.~\cite{castellanos-gomez_why_2016, RevExTMDC} and references therein.
Besides the four main representative TMDCs MoS$_2$, MoSe$_2$, WS$_2$, and WSe$_2$
many further materials have been identified theoretically
owning various interesting optical properties \cite{c2db}.

\begin{figure}[t]
  \centering
  \includegraphics[width=.95\linewidth]{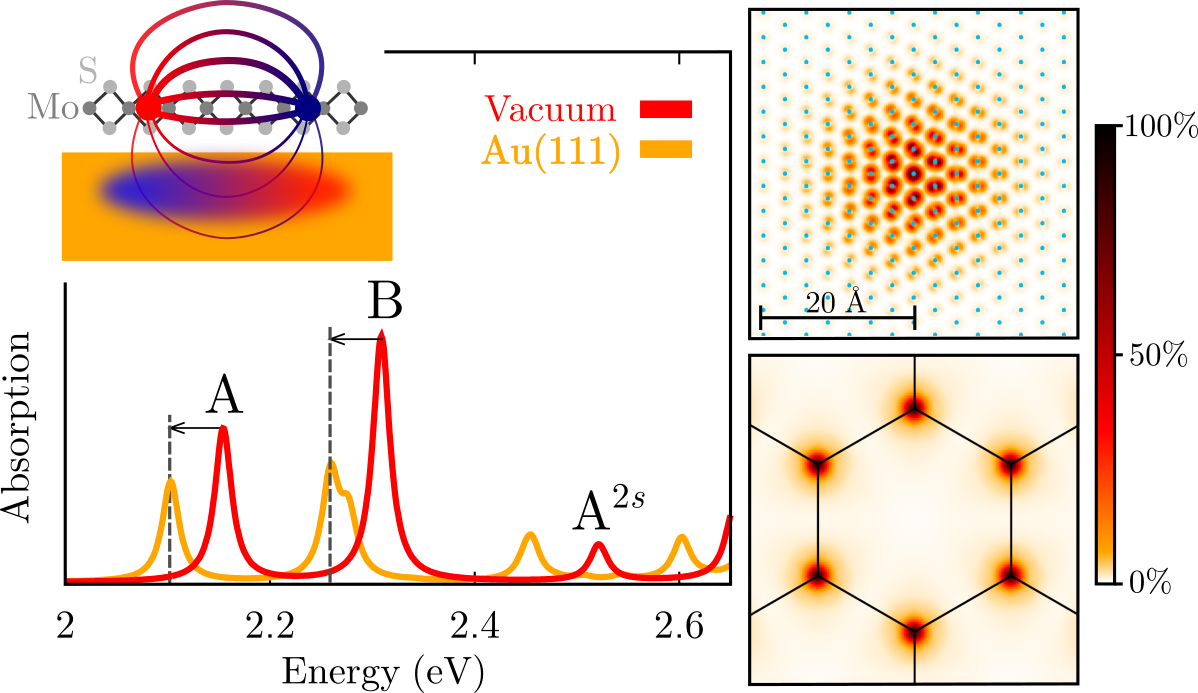}
  \caption{%
    Absorption of an MoS$_2$ monolayer in vacuum (red) and on a substrate (Au(111), yellow).
    The inset schematically depicts the field-lines of an exciton above a substrate
    owning a strongly screened electron-hole interaction.
    In the right top panel the modulus squared of the vacuum exciton wave function of the A exciton
    in real space is shown on the plane spanned by the Mo atoms (the hole is fixed close to the central Mo atom).
    The lower right panel shows the corresponding modulus of the contribution to the exciton in $k$-space.
    The color from while to white to black denote the strength of the excitation.
    Modified from \cite{MoS2Dru}.
  }
  \label{fig:MoS2}
\end{figure}
In Fig.~\ref{fig:MoS2} the prototypical calculated absorption spectrum of MoS$_2$ is shown.
The two most prominent A and B excitons stem from transitions close to the
direct gap at K \cite{Qiu.2013}
and their splitting results from the spin-orbit splitting of the involved valence bands.
Interestingly, in TMDCs valleys and spins are coupled \cite{DiXiao.2012, Cao.2012}
which allows for specific excitations using circular polarized light.

The reduced and anisotropic screening of two-dimensional monolayers have several
interesting consequences.
Typical exciton binding energies of about $0.8$\,eV (for MoS$_2$)
and up to about $2.0$\,eV for wide band-gap materials like $h$BN
have been calculated \cite{wirtz_excitons_2006, ExiplexIL}.
At the same time the higher excited state
experience spatially different dielectric screening
which results in strong deviations
from a Rydberg-like series \cite{qiu_screening_2016}.
A further consequence of the two-dimensionality
is the vulnerability to environmental screening \cite{MoS2Dru}.
Fig.~\ref{fig:MoS2} shows the modification of the absorption spectrum
of MoS$_2$ placed on a gold substrate.
While the band gap is reduced significantly by $0.6$\,eV
the electron-hole interaction is lowered
by nearly the same amount and
results in a red-shift of $0.05$\,eV (for the A exciton) only.
Consequently, in heterostructures
the neighbouring layer can be utilized to
fine-tune the properties of intralayer excitons.

\subsection{Interlayer excitons}\label{theo:inter}
In stacked two-dimensional materials,
like bilayers, trilayers, etc. up to bulk,
the neighbouring layers interact with each other.
This interaction, even if it is typically weak, results in a hybridization
and leads to excitons which are localized on different layers.
This concept of interlayer excitons (IX) has been utilized previously for double-well systems,
and similarly for charge transfer excitations (CT) in molecular systems
\cite{Danan_1987, Jacquemin_2017}.

As soon as two subsystems (i.e. layers) start to interact, interlayer excitons can and will form.
However, due to the spatial separation of electron and hole on different layers
the exciton binding energy of these states are typically smaller,
e.g. in a hetero-bilayer of TMDCs the interlayer exciton binding energy
has been calculated to about $0.35$\,eV compared to about $0.5$-$0.6$\,eV for the intralayer states \cite{Deilmann.2018, Torun_2018}.
Furthermore, the oscillator strength is much weaker (compared to bright intralayer states).
This is why interlayer excitons have been overlooked for some time.
\begin{figure}[t]
  \centering
  \includegraphics[width=.95\linewidth]{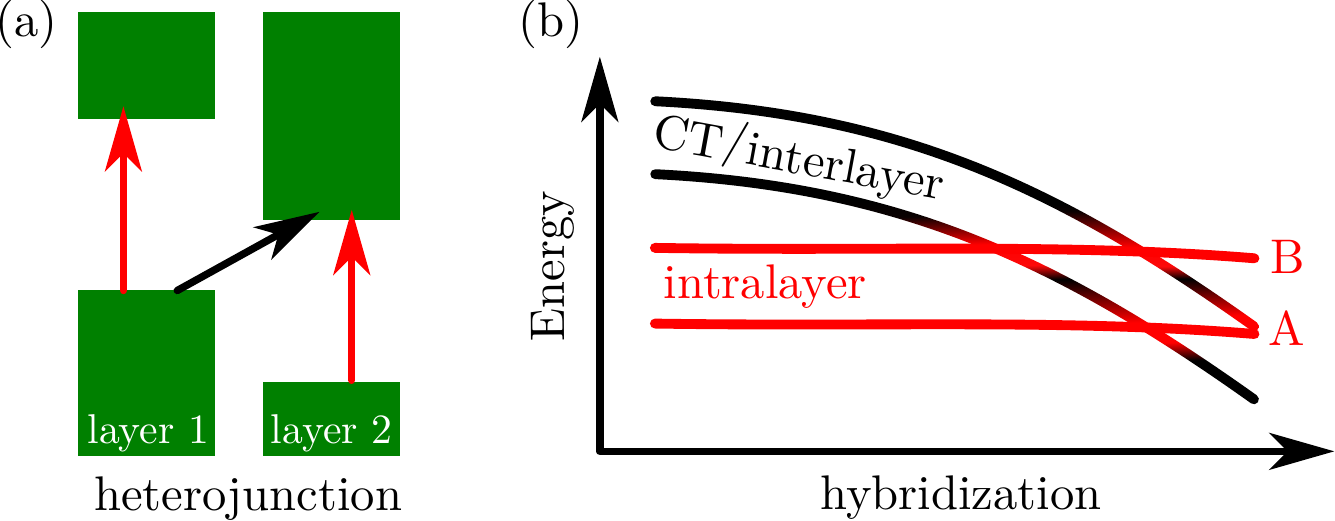}
  \caption{%
    Schematic sketch of intralayer (red) and interlayer excitons (black).
    (a) In a heterojunction (type II) interlayer excitons
    can have a lower excitation energy even though the exciton binding energy is reduced
    and become the excitonic ground state.
    (b) By varying the hybridization of neighbouring layers
    the interlayer states can strongly hybridize and
    gain large oscillator strength (marked as red color) \cite{mixedIL,Alexeev.2019}.
  }
  \label{fig:IL}%
\end{figure}
However, in two different situations interlayer states gain a curtail role (Fig.~\ref{fig:IL}).
On the one hand, the neighbouring layer may have a band offset
which leads to a type II heterojunction.
In this case the interlayer exciton can become the energetically lowest state
and is thus highlighted especially in photoluminescence.
On the other hand,
if the energy is close to those of an intralayer exciton these states can hybridize
and lead to mixed IX excitons with large oscillator strength.

In recent years, especially for TMDC heterostructures several studies have discussed interlayer states
\cite{latini_interlayer_2016, MoTe2bulk, Gao.2017, Nayak.2017, Deilmann.2018, Gillen_2018, mixedIL, Torun_2018, MoWbulkIL, ExiplexIL, Kunstmann.2018, Gerber_2019}.
The considered systems range form hetero-bilayers
(such as MoS$_2$/WS$_2$ or MoSe$_2$/WSe$_2$),
homobilayers (such as MoS$_2$/MoS$_2$),
to bulk systems (e.g. MoTe$_2$).
While in several hetero-bilayers (type-II)
interlayer excitons have been identified as lowest energy state,
in homobilayers and bulk interlayer states
with strong oscillator strengths have been found at higher energies.

\section{Multivalley Physics of Interlayer Excitons}\label{sec:multi}

In a simplified picture, interlayer excitons in hetero-bilayers are created by charge transfer across the vdW interface followed by the formation of excitonic bound states with the electron states fully localized in one layer and the hole state fully localized in the other layer \cite{Rivera.2015, Wurstbauer.2017}. Localization of electronic states in adjacent layers result in a significantly reduced overlap of the electron and hole wave-function and a sizable permanent out-of plane dipole moment \cite{Ciarrocchi.2018}. This picture is supported by calculations of the band offsets of the TMDC monolayers that predict a type II band alignment, as sketched in \ref{fig: 1}b, for any combination of two different semiconducting TMDCs \cite{Kang.2013, Liang.2013, Ozcelik.2016}. Calculations including interlayer interaction confirm the band alignment considering the $K$ points, but predict energy shifts of the valence band close to the $\Gamma$ point and of the conduction band around the $\Sigma$ point similar to the difference of the band structures of mono- and bilayer crystals of TMDCs \cite{Kosmider.2013, Miller.2017}. The location and nature of the smallest energy gap in TMDC hetero-bilayers therefore depends on the considered material combination, doping, dielectric environment. In a precise description, the hetero-bilayers are not considered in a model consisting of two independent monolayers \cite{Wang.2017}, but as a newly formed van der Waals solid taking into account hybridization of electronic states. At the $K$ valleys, interlayer coupling is small \cite{Wang.2017c} and hybridization is not found to have a large affect at the $K$ valleys \cite{Kosmider.2013}. The changes of the valence band at the $\Gamma$ valley and the conduction band at the $\Sigma$ valley due to interlayer coupling are of the same size as the band offsets of the hetero-bilayers \cite{Wang.2016}, resulting in strong mixing of the electronic states in these valleys \cite{Kosmider.2013, Gao.2017}. Experimental results obtained by scanning tunneling spectroscopy, X-ray photoemission spectroscopy and angle resolved photoemission spectroscopy (ARPES) provide insights on the electronic bands of different types of vdW hetero-structures and corroborate type-II band alignment and the hybridization of electronic states at the $\Gamma$ and $\Sigma$ points \cite{Chiu.2015, Hill.2015, Wilson.2017}. ARPES measurements indicate that the topmost valence band states for a \mole/\wole\ heterostructure  is still located at the $K$ point \cite{Wilson.2017}. In photoluminescence experiments, signatures for the formation of IX in TMDC hetero-bilayers have been observed for all possible combination of the four most prominent semiconducting TMDCs, \moly, \woly, \mole, and \wole\ \cite{Ceballos.2014, Rivera.2015, Jin.2019, Gong.2014, Fang.2014, Mouri.2017}.

\begin{figure}
\includegraphics[scale=0.7]{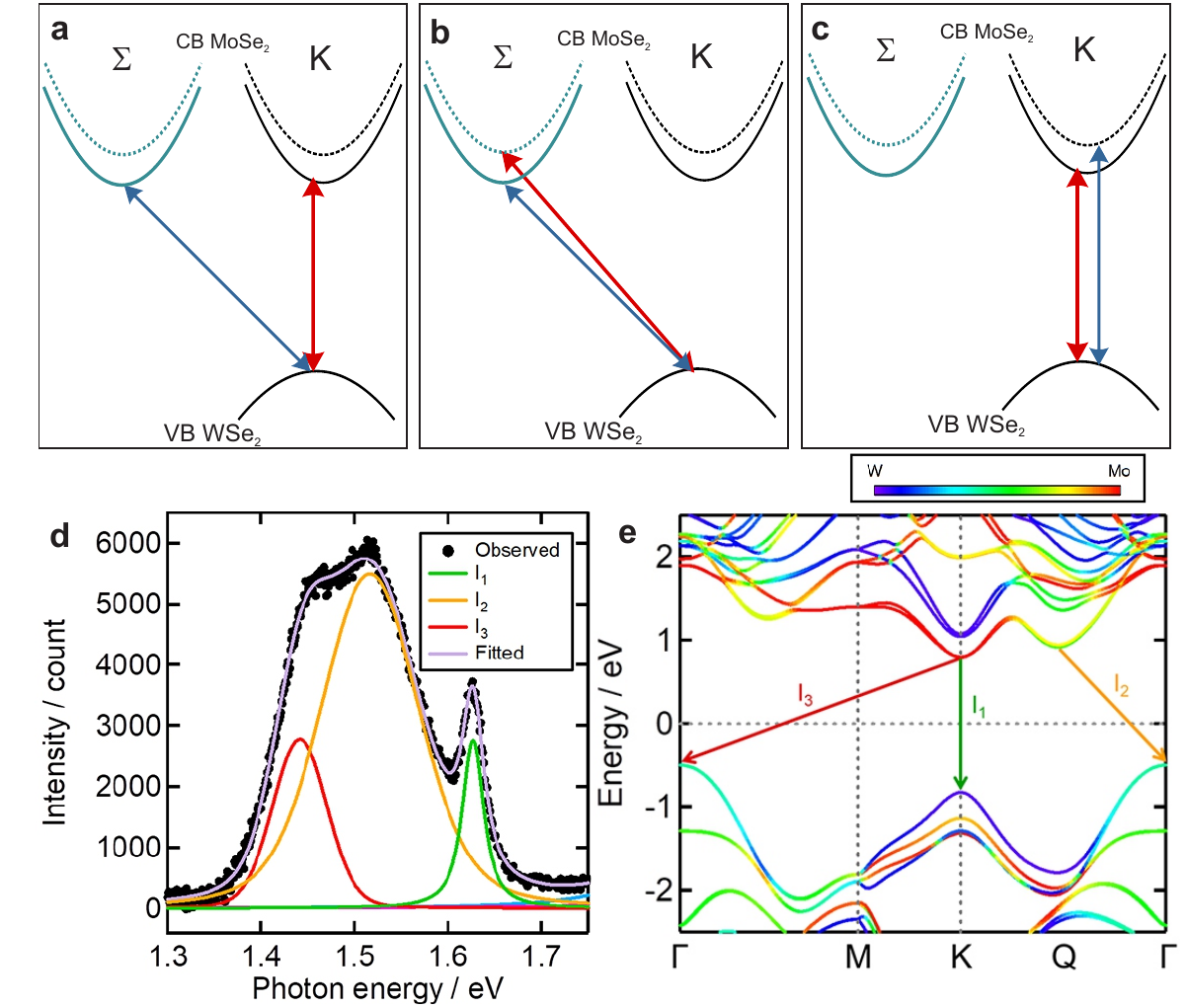}
\caption{Different interpretations for the observed splitting in IX emission from a \mole/\wole\ hetero-bilayer:
(a) Momentum direct and momentum indirect IX involving $\Sigma$ and $K$ valleys \cite{Miller.2017}.
(b) Momentum indirect IX formed between spin-orbit split conduction band states at the $\Sigma$ valley with valence band states at the $K$ valley \cite{Hanbicki.2018}.
(c) Momentum direct IX formed between spin-orbit split conduction band states at $K$ valley with valnece band states at the $K$ valley \cite{Ciarrocchi.2018}.
(d) Room temperature IX photoluminescence spectrum of a hBN encapsulated \moly/\woly\ hetero-bilayer clearly indicating a triplet structure. The individual contributions I$_{1}$, I$_{2}$, I$_{3}$ are included from Voigt fits to the spectrum \cite{Okada.2018} (similar spectra shown in \cite{Kiemle.2019}).
(e) DFT band structure of \moly/\woly\ hetero-bilayer with projections of bands onto individual layers are shown with a color gradient showing high degree of hybridization at the $\Gamma$ and $Q/\Sigma$ point \cite{Okada.2018}.
[(d,e) Reprinted with permission from \cite{Okada.2018}.]
}
\label{fig: 5}
\end{figure}

The lowest conduction bands and topmost valence bands of a \moly/\woly\ hetero-bilayer from ab initio calculation based on density functional theory and many-body perturbation theory including the contributions of the atomic orbitals of the individual layers to the hybridized bands are shown in figure \ref{fig: 1}d. The mixing of orbital states is particularly strong at the $\Gamma$ point of the valence band and the $\Sigma$ point (also referred to as $Q$ or $\Lambda$ point) of the conduction band. The hybridization shifts the global maximum of the valence band from the $K$ point to the $\Gamma$ point and the global minimum of the conduction band from the $K$ point to the $\Sigma$ point \cite{Kosmider.2013, Gillen_2018, Deilmann.2018, Gao.2017, Kiemle.2019}. In general, the existence of multiple valleys that are very close in energy demands for a careful determination of the excitonic states contributing to IX formation in TMDC hetero-bilayers. Different degrees of hybridization of the electronic states of the constituent layers can cause the charge carriers to be delocalized across the hetero-bilayer at specific points in the Brillouin zone, while the carriers at other points remain nearly completely localized in one of the layers \cite{Gao.2017, Okada.2018, Gillen_2018, Deilmann.2018, Torun.2018}. In this regard, such structures have to be thought in terms of `novel artificial vdW solids', more than just a sum of two 2D materials. Exact energies of the valley specific electronic states as well as hybridization of bands have been reported to depend crucially on the material combination, their rotational alignment and their dielectric environment\cite{Miller.2017, Kunstmann.2018, Nayak.2017}. The crystal alignment not only determines the energy of the electronic bands, but also the location of the different valleys in momentum space \cite{Yu.2015}.

In photoluminescence experiments, multiplet emission lines from IX recombination are reported from different combinations of TMDC hetero-structures, pointing towards a complex interplay between multi-valley physics and hybridization of electronic states \cite{Miller.2017, Hanbicki.2018, Baranowski.2017, Kunstmann.2018, Ciarrocchi.2018, Calman.2019, Kiemle.2019}. Doublet structures have been observed in the photoluminescence from \mole/\wole\ hetero-bilayer by several groups \cite{Miller.2019, Hanbicki.2018, Ciarrocchi.2018, Calman.2019}. From temperature and excitation power dependent experiments together with DFT calculations of the bandstructure, Miller \textit{et al.} interpret the emission doublet as a momentum indirect transition between $\Sigma$ and $K$ valley and a momentum direct interlayer transition between the $K$ valley that are separated by about 40\,meV \cite{Miller.2017} as sketched in figure~\ref{fig: 5}a. This combination of \mole\ and \wole\ TMDC monolayers results in a momentum indirect lowest energy transition between $\Sigma^{\text{MoSe}_2}$ and $K^{\text{WSe}_2}$ that is activated by an $M$ point phonon for momentum conservation. This indirect optical interband transition is most likely mediated by the longitudinal acoustic mode at the $M$ point LA(M) featuring a high electron-phonon coupling strength \cite{Ge.2013}. The energetically slightly higher lying transition between the conduction band state $K^{\text{MoSe}_2}$ and the valence band state $K^{\text{WSe}_2}$ is slightly indirect in momentum space due to residual rotational and lattice mismatch. Radiative decay for IX located at the $K$ points is possible for nearly aligned crystal axes as long as the IX kinematic momentum is sufficiently large to meet the light cone \cite{Rivera.2015, Yu.2015, Miller.2017}. 
Photoexcitation introduces and lifts a quasi Fermi level for electrons and holes.
Hence both charge carriers gain quasi Fermi momenta and their sum can become sufficient to compensate for the displacement vector between $K^{\text{MoSe}_2}$ and $K^{\text{WSe}_2}$. This interpretation holds for \mole/\wole\ heterostructures with nearly AB (60 degrees) of nearly AA (0 degreed) stacking order \cite{Miller.2017}. Baranowski \textit{et al.} used a very similar interpretation to explain the experimentally observed doublet structure in the photoluminescence from IX hosted by \moly/\mole/\moly\ tri-layer structure \cite{Baranowski.2017}. A reduced splitting of the IX emission lines of about 25~meV for \mole/\wole\ hetero-bilayers is reported by Calman \textit{et al.} \cite{Calman.2019}, by Hanbicki \textit{et al.} \cite{Hanbicki.2018}, and Ciarrocchi \textit{et al.} \cite{Ciarrocchi.2018}, respectively, with contrary interpretations. Calman \textit{et al.} explain the observed splitting due to neutral and charged interlayer excitons \cite{Calman.2019} in agreement with the interpretation for a IX doublet in a \wole/\mole/\wole\ structure \cite{Choi.2018}.
Whereas Hanbicki \textit{et al.} interpret the IX doublet in terms of momentum indirect interlayer transitions between electron states hosted in the spin split conduction band at the $\Sigma$ valley with hole states localized at the $K^{\text{WSe}_2}$ valley (see figure \ref{fig: 5}b) \cite{Hanbicki.2018}. The energy difference of 25\,mev fits to the splitting due to spin-orbit coupling of the hybridized electronic states at the $\Sigma$ valley \cite{Hanbicki.2018}. The authors point out that the transitions are not purely interlayer in character since the wavefunction of the electronic state at the $K$ point have significant weight in both layers, while the hole state is localized in \wole\ \cite{Hanbicki.2018}. Similarly, Ciarrocchi \textit{et al.} attribute the doublet structure to a momentum direct transition from the spin orbit split conduction bands at $K^{\text{MoSe}_2}$ to the topmost valence band state at $K^{\text{WSe}_2}$ \cite{Ciarrocchi.2018} as depicted in figure \ref{fig: 5}c. Another group reports momentum direct ($K-K$) and momentum indirect ($K-\Gamma$) IXs in a \moly/\wole\ hetero-bilayer with the emission energy and intensity strongly dependent on the twist angle of the adjacent layers \cite{Kunstmann.2018}. Also this groups points towards the effect of hybridized electronic states since the momentum indirect IX are interpreted to be formed by an electron state localized in \moly\ at the $K^{\text{MoS}_2}$ point and delocalized holes in the hybridized $\Gamma$ point \cite{Kunstmann.2018}. As demonstrated in figure \ref{fig: 5}d, even triplet IX emission has been reported for hBN encapsulated \moly/\woly\ hetero-bilayers and interpreted as momentum direct and indirect IX transitions involving electron states at $K$ and the hybridized $\Sigma$ valley and hole states at $K$ and the hybridized $\Gamma$ valley indicated in figure \ref{fig: 5}e \cite{Okada.2018, Gao.2017, Deilmann.2018, Gillen_2018, Kiemle.2019}.
Finite coupling of electronic states and valley selective hybridization together with the finite dipole moment of IX allows to tune and engineer the properties of the bound exciton states by application of an electric field normal to the hetero-bilayers utilized by gate electrodes \cite{Gao.2017, Kiemle.2019, Lucatto.2019}. Electric field control of layer index, orbital character, lifetime and emission energy of indirect excitons in \HS\ hetero-bilayers embedded in an vdW field effected structure allows for the design of novel vdW based quantum-nano architectures as e.g. the suggested realization of charge qubits in a vdW hetero-bilayer \cite{Lucatto.2019}.

\section{Crystal alignment and moiré excitons}\label{sec:moire}

\begin{figure}
\includegraphics[scale=1]{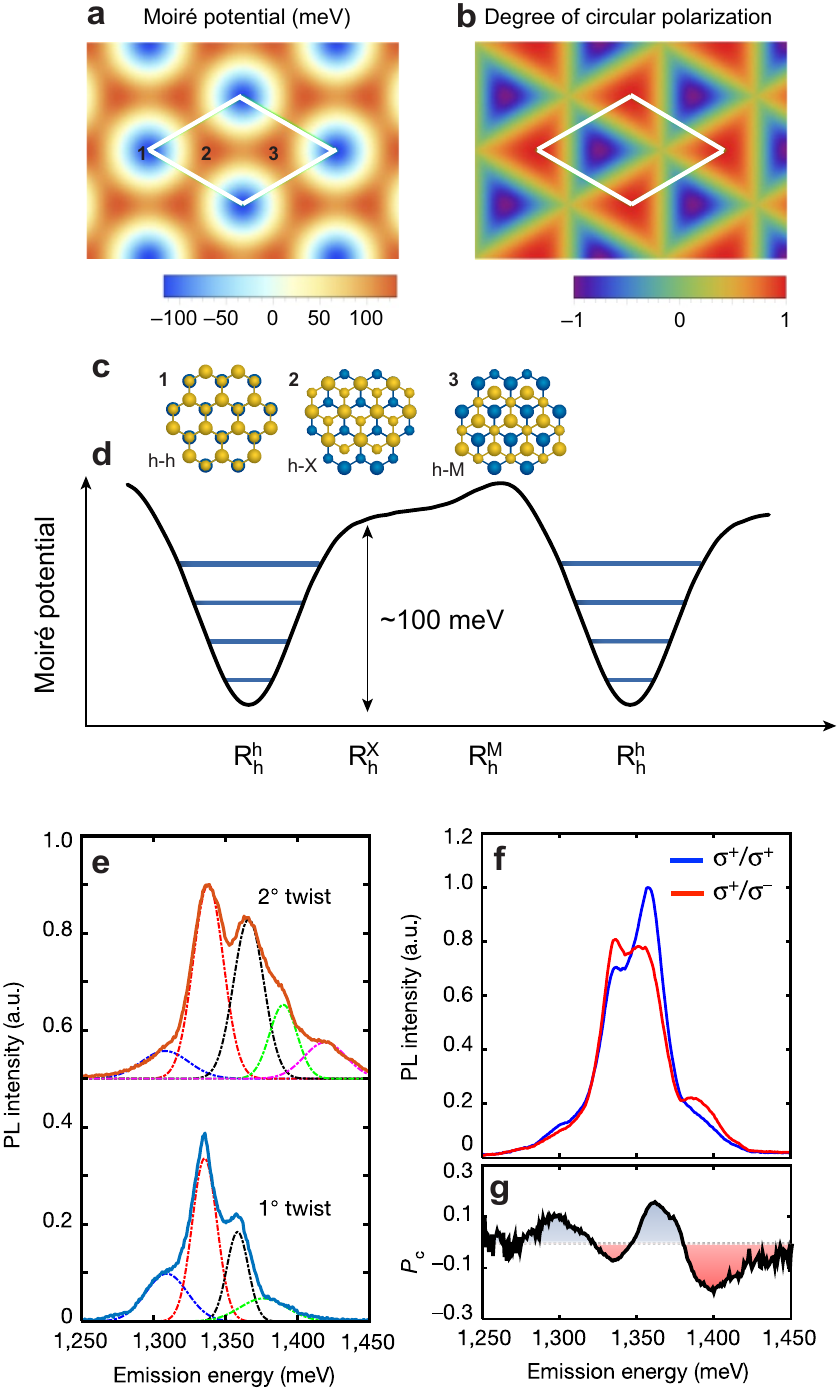}
\caption{Moiré superlattice potential and localized moiré excitons in a \mole/\wole hetero-bilayer \cite{Tran.2019}.
(a) Moiré potential of the interlayer exciton transition showing a local minimum at the h-h site (1) and local maxima at the h-X (2) and the h-M (3).
(b) Spatial map of the optical selection rules for $K$-valley excitons. The high-symmetry points of the moiré potential are circularly polarized and the regions in between are elliptically polarized.
(c) Sketch of the atomic registry for the stacked layers at high symmetry points (1), (2) and (3) of the moiré potential as indicated in (a).
(d) Spatial variation of the moiré potential and the confined multiple interlayer exciton resonances.
(e) Photoluminescence spectra from for samples with twist angles of $1\deg$ (bottom) and $2\deg$ (top). The spectra are fitted with four (1$\deg$) and five (2$\deg$) Gaussian functions, respectively.
(f) Circularly polarized IX emission spectra of the 1$\deg$ sample.
(g) Degree of circular polarization for the spectra from (f).
[Panel a, b, d-g reprinted with permission from \cite{Tran.2019}].
}
\label{fig: 6}
\end{figure}

Because of the weak van der Waals forces between adjacent layers in a heterostructure, the TMDC monolayers keep their individual in-plane lattice constants $a$ \cite{Kang.2013b}, which vary by 0.15-4\% between \moly\ ($a=3.160$\,\AA) , \woly\ ($a=3.155$\,\AA), \mole\ ($a=3.288$\,\AA), and \wole\ ($a=3.280$\,\AA) \cite{Wildervanck.1964, Brixen.1962}.
Strain-free vdW assembly of hetero-stacks implies a spatial alternation of the atomic registry,
i.e. the formation of a moiré pattern \cite{Kang.2013b}.
As depicted in figure~\ref{fig: 1}e an additional twist angle between the crystal axes of the individual layers can further tune the moiré structure.

Most real structures exhibit a finite mismatch of the crystal axes. Deviation from the atomic registry of the 2H stacking (AB stacking) of natural homo-bilayers changes the interlayer distance due to steric effects \cite{Liu.2014,Kang.2013b, Lu.2017}. The interlayer distance is smallest, but not equal, for the AA and AB stacking corresponding to a twist angle between adjacent layers of $0\deg$ and $60\deg$, respectively \cite{Lu.2017}. The interlayer distance sensitively influences the electronic bands \cite{Liu.2014,Kang.2013b, Lu.2017}. The most pronounced changes occur, where hybridization between the two layers is strong, that is in particular at the $\Gamma$ valley of the valence band \cite{Liu.2014, Kang.2013b}. An increase of the absolute energy of the valence band at the $\Gamma$ point of up to 120 meV for twist angles that deviate significantly from $0\deg$ or $60\deg$ in artificially stacked \moly\ homo-bilayers has been determines from ARPES \cite{Yeh.2016}. This trend is reflected in the twist angle dependence of the transition energy, the intensity and the decay time of interlayer excitons in hetero-bilayers \cite{Liu.2014, Nayak.2017, Kunstmann.2018, Heo.2015}. In addition, slightly modified IX transition energies for AA and AB stacking order in hetero-bilayers have been reported theoretically \cite{Lu.2017} and experimentally \cite{Hsu.2018}.  

The charge transfer rate of photo-excited electron hole pairs is determined to be independent of the twist angle \cite{Wang.2016, Zhu.2017}. This is explained by a  much larger band offsets compared to dependence of the band energies on the twist angle \cite{Wang.2017}. In contrast, the formation and cooling dynamics of IX is dependent on the twist angle between adjacent TMDC monolayers \cite{Merkl.2019}.\par
For stacked TMDC homo- and hetero-bilayers, the atomic registry depends on the lateral arrangement of the two layers. For AA and AB stacking order the hollow sites of the two hexagonal lattices of the individual monolayers are aligned to each other. For this configuration, the interlayer distance is highest compared to an alignment of the hollow site to the metal or the chalcogene site \cite{Zhang.2017,Yu.2017}. The combined mismatch of the lattice constants and/or the rotational alignemnt (twist) between adjacent monolayers results in a periodic variation of the local atomic registry forming an hexagonal moiré superlattice pattern with a length scale in the order of 10 nm. The size of the moiré super-cell (figure \ref{fig: 1}e) depends on the lattice mismatch and rotational alignment and gets smaller with increasing twist, as a general tendency. The threefold rotation symmetry of the local atomic registry is restored at three points within the super-cell highlighted with circles in figure \ref{fig: 1}e and explicitly shown in figure \ref{fig: 6}a-c. These high symmetry points are local energy extreme within the moiré supercell (figure \ref{fig: 6}a, d). The result is a periodic modulation of the energy landscape in real space with a magnitude in the order of up to 0.15 eV \cite{Zhang.2017, Tran.2019}. The periodically modulated energy landscape in suitable TMDC hetero-bilayers enables the formation of spatially localized IX states trapped at different superlattice sites \cite{Seyler.2019, Tran.2019, Jin.2019} with moiré quasi angular momentum periodically switching the optical selection rules \cite{Jin.2019b} as demonstrated in figure \ref{fig: 6}b,d-g.

The experimental signatures of those trapped moiré IX appearing just below the direct exciton emission are an ultra-narrow line-width in the order of 100\,$\mu$eV, varying optical selection rule and a peculiar dependence on rotational alignment of the hetero-bilayer \cite{Tran.2019, Jin.2019, Seyler.2019}. In particular, Tran \textit{et al.} observed multiple IX resonances with alternating circularly polarized emission in a hBN encapsulated \mole/\wole\ hetero-bilayer with small twist \cite{Tran.2019}. The resonances are interpreted as ground state, excited states and phonon replica of IX confined within potential minima of the moiré potential \cite{Tran.2019} as demonstrated in figure \ref{fig: 5}. Similarly, Seyler \textit{et al.} report experimental evidence for ultra narrow (linewidth of about 100\,$\mu$eV) photoluminescence from interlayer valley excitons that are trapped in the moiré potential of a \mole/\wole\ with low twist angle \cite{Seyler.2019}. Most significantly, in magneto-photoluminescence experiments the authors find that the $g$-factor only takes two values across a sample that matches those of interlayer excitons with -15.9 and 6.7, respectively \cite{Seyler.2019}. From the $g$-factor results together with the circularly polarized emission strongly suggesting three-fold rotation symmetry of the emission sites, the author conclude the existence of trapped moiré excitons even in the presence of inhomogeneities and sample non-uniformity resulting in multiple emission line with similar characteristics \cite{Seyler.2019}. Also for a different heterostructure, hBN encapsulated closely aligned \wole/\woly\ hetero-bilayer, Jin \textit{et al.} report the observation of moiré superlattice exciton states emerging as multiple emission peaks slightly red-shifted from the neutral \wole\ intralayer exciton \cite{Jin.2019}. Interestingly, the authors observe strong resonances that show a clear doping dependence, not only in emission but also in absorption spectra \cite{Jin.2019}. The moiré potential does not only affect the single particle energies in twisted TMDC homo- and hetero-bilayers, but also induced moiré exciton flat bands \cite{Wu.2017, Yu.2017, Wu.2018, Jin.2019}. Strong interlayer interactions between the adjacent \wole\ and \woly\ mono-layers are required to enter this strong coupling regime for excitons in which the moiré excitons are spatially trapped at well separated periodic sites in the moiré potential landscape \cite{Jin.2019}. Those moiré excitons in \wole/\woly\ are described in a theoretical framework in which the exciton kinetic energy is much lower than the moiré potential generating multiple flat exciton minibands \cite{Jin.2019}. Together with the very narrow excitonic emission lines, the flat exciton bands are promising for the realization of topological excitons and to explore a correlated exciton Hubbard model \cite{Jin.2019}.

More general, moiré superlattices in nearly aligned TMDC homo- and hetero-bilayers provide a promising platform to realize and study Hubbard model physics in a new widely tunable solid state platform \cite{Wu.2018}. In partially filled flat moiré bands strong correlations may result in exciting many-body ground states including spin-liquid states, quantum anomalous Hall insulators, chiral $d$-wave superconductors \cite{Wu.2018}, and for exactly half filling a Mott-insulator state \cite{Tang.2019}.

\section{Conclusion and Future Perspectives}\label{sec:out}
The interaction of two-dimensional materials
and their van der Waals heterostructures with light
is a versatile and fast-growing field.
Due to their unique properties the class of transition metal dichalcogenides
is still mostly used, even though further novel materials are proposed repeatedly.
A central step is the synthesis
(direct growth by chemical vapor deposition or molecular beam epitaxy,
liquid exfoliation, or micromechanical exfoliation)
and the reproducible stacking of the high-quality vdW materials.
The rotational alignment of the crystal axes during the transfer process
is possible by optical microscopy because of their preferable breaking points
and can be effectively checked by polarization resolved second harmonic generation.
The atomically resolved structure of heterostructures results in moiré patterns
with a curvature as well as spatially varying distances of the individual layers.
An accurate theoretical description from first principles
requires many-body perturbation theory (most often in the $GW$/BSE approximation).
However, its applicability is limited to a few atoms only
and for the description of huge moiré patterns simplified models have to be employed.

Even though the properties of the individual layers mostly govern the interaction with light,
the interlayer mixing is able to fine-tune these properties,
but may also lead to completely new phenomena.
Therefore, such van der Waals solids crucially depend
on the spatially varying coupling
which leads to two prominent effects:
The hybridization of different valleys in the band structure changes
and interlayer excitons, i.e. electron-hole pairs with different amounts of electrons and holes on different layers.

Due to the effective charge separation in such interlayer excitons
several interesting properties can be designed
which may be interesting for instance for quantum-technology devices.
Interlayer excitons own long lifetimes which allows for thermalization and diffusion over several micrometers.
Furthermore, the out of plane dipole can be easily manipulation by electric fields
and promotes van der Waals solids
as light-emitting diodes, energy conversion, opto- and valleytronics, as well as for exciton based information technologies.

Moreover, vdW structures are also interesting candidates to study fundamental science.
As recently demonstrated, particular designed WSe$_{2}$/MoSe$_{2}$heterostructures have the potential
to investigate collective excitonic phenomena
like high-temperature superfluidity and Bose-Einstein condensation \cite{Wang.2019, Sigl.2020}.
Very recently several works have started to investigate
novel properties of moiré structures and their interaction with light.
Tang \emph{et al.} \cite{Tang.2019} have proposed the
WSe$_2$/WS$_2$ heterostructure to probe the Hubbard model.
Also in WSe$_2$/WS$_2$ Regan \emph{et al.} \cite{1910.09047}
have observed several unexpected insulating states.
Wang \emph{et al.} \cite{1910.12147} have
investigated twisted bilayer WSe$_2$.
For twist angles spanning from $4\deg$ to $5.1\deg$
the authors observe a flat band
which has been put into context of superconductivity in graphene.
Considering the quite rapid developments in this area in the last years,
it is very likely that many extremely exciting developments will continue in near future.

\section*{Acknowledgements}
We thank Bastian Miller, Florian Sigger, Jonas Kiemle and Fabian Kronowetter for support with the figure preparation.
We acknowledge financial support from the German Research Foundation (Germany's Excellence Strategy – EXC 2089/1 – 390776260, excellence cluster `Nanosystems Initiative Munich' (NIM) and DFG Projects No. DE 2749/2-1 and WU 637/4-1).

\end{document}